\documentclass{aa} \usepackage{graphics}
 
\begin{document}
 
\thesaurus{07 
				  (07.13.1; 
				   07.13.2)} 
 
\title{On the airburst of large meteoroids in the Earth's atmosphere} 
\subtitle{The Lugo bolide: reanalysis of a case study}
 
\author{Luigi Foschini}
 
\offprints{L. Foschini}
 
\institute{CNR - Institute FISBAT, Via Gobetti 101, I-40129 Bologna, 
Italy; (email: L.Foschini@fisbat.bo.cnr.it)}
 
\date{Received March 31, 1998; accepted July 14, 1998}
 
\titlerunning{On the airburst of large meteoroids\ldots} 
\authorrunning{L. Foschini}
 
\maketitle

\begin{abstract}
On January 19, 1993, a very bright bolide (peak magnitude $-23$) 
crossed the sky of Northern Italy, ending with an explosion 
approximately over the town of Lugo (Emilia Romagna, Italy).  The 
explosion (14 kton of energy) generated shock waves which were 
recorded by six local seismic stations.  A reanalysis of the available 
data leads us to the hypothesis that the meteoroid was a porous 
carbonaceous chondrite, somehow similar in constitution to the 
asteroid 253 Mathilde.  
\end{abstract}
 
\section{Introduction}
The atmospheric interaction of large meteoroids provides our primary 
tool to characterize their population, physical and chemical 
properties, and dynamical evolution.  In turn, this can lead to a 
better understanding of the diverse populations of small bodies of the 
Solar System.  Currently, our knowledge is still quite limited, 
although, especially after the impact of comet D/Shoemaker--Levy~9 on 
Jupiter, the research efforts in this field have been intensified.  In 
particular, in 1994 the US Department of Defense made of public domain 
its records on energetic bolides over a time span of about twenty 
years (Tagliaferri et al.  \cite{DOD}).  These data indicate that, 
from 1975 to 1992, there were 136 airbursts of energy greater than 1 
kton, but the real number was probably at least 10 times higher, 
because the satellite system does not cover the entire Earth surface.
 
Both data and theories are required to assess the impact hazard and to 
understand the very bright bolides.  From this point of view, the Lugo 
bolide is a very interesting event, because the airburst was detected 
by several seismic stations.  The corresponding data allow us to 
characterize the meteoroid and to draw some tentative inferences about 
its nature and origin.  We have carried out a reanalysis of this event 
and we found that the data are most consistent with the hypothesis 
that the involved meteoroid was a porous carbonaceous chondrite, 
somehow similar to the asteroid 253 Mathilde.
 
\section{The Lugo bolide}
On 1993 January 19 at 00:33:29 UT a large meteoroid entered the 
atmosphere approximately over the town of Lugo, in Emilia Romagna, 
Italy.  The impact was recorded by the National Research Council (CNR) 
forward--scatter meteor radar and by six seismic stations, three 
belonging to the Microseismic Network of Ferrara (Pontisette, C\`{a} 
Fornasina, Fiorile d'Albero) and the others to the National Institute 
of Geophysics (Barisano, Santa Sofia, Poggio Sodo).  The event was 
also observed by several eyewitnesses, as it lit an extremely large 
area (almost all of Italy), and they reported a visual magnitude in 
the range $-22$ to $-25$.  Preliminary calculations were carried out 
based on the eyewitness reports, although they were fragmentary and 
sometimes contradictory (Cevolani et al.  \cite{LUGO1}, Korlevi\'{c} 
\cite{LUGO3}).  Only at a later time we found seismic data which 
enabled us to infer the location of the explosion (Cevolani et al.  
\cite{LUGO2}).  This analysis indicated that a meteoroid of initial 
radius in the range $1.5\div 3$~m impacted the Earth atmosphere at a 
velocity of about $26$~km/s, with an inclination of the trajectory to 
the horizon of $8\degr\div 20\degr$.  By means of the seismic data, it 
was possible to calculate the height ($30\pm 3$~km), latitude 
($44{\degr}.48\pm 0{\degr}.01$~N) and longitude ($11{\degr}.91\pm 
0{\degr}.01$~E) of the explosion.
 
\section{The reanalysis: Aerodynamics}
Here, we will assume that the only reliable data are those recorded by 
the seismic stations, which in general are a very useful tool for 
understanding this kind of airburst (e.g.  Ben-Menahem 
\cite{TUNGUSKA1}).  Therefore, we assume as valid the height, latitude 
and longitude of the explosion only, i.e.  those data calculated from 
seismic data (Cevolani et al.  \cite{LUGO2}).

The aerodynamics of large meteoroid/small asteroid impacts has been 
studied by several authors, sometimes with special reference to the 
1908 Tunguska explosion (e.g.  Ceplecha and McCrosky \cite{PE}, 
Ceplecha et al.  \cite{CEPLECHA1}, Chyba et al.  \cite{TUNGUSKA2}, 
Hills and Goda \cite{HILLS}, Lyne et al.  \cite{TUNGUSKA3}).  Although 
the details may vary, there is a consensus that a 30~km explosion 
height is typical for a carbonaceous chondrite or a cometary body.  In 
the theory of Hills and Goda (\cite{HILLS}) the height of first 
fragmentation is calculated comparing the stagnation pressure in front 
of the meteoroid ($P_{max}=\rho_{0}V_{e}^{2}$) to the mechanical 
strength $S$ of the cosmic body.  We rearrange the formula to evaluate 
the meteoroid speed ($V_{e}$):

\begin{equation}
		V_{e}=\sqrt{\frac{S}{\rho_{0}}\exp\left[\frac{h_{e}}{H}\right]}\
		\label{e:velo}
\end{equation}

\noindent where $\rho_{0}$ is the atmospheric density at the sea level 
[kg/m$^{3}$], $h_{e}$ is the height of first fragmentation [km] and $H$ is the 
atmospheric scale height (about 8~km).  For the strength, we assume 
$S=10^{7}$~Pa, that is an intermediate value between those appropriate 
for carbonaceous chondrites and for cometary bodies.  We obtain 
$V_{e}=18\pm 3$~km/s, that is a value much lower than that derived 
previously (about 26~km/s).
 
\begin{figure}[t]
		\centering
		\includegraphics{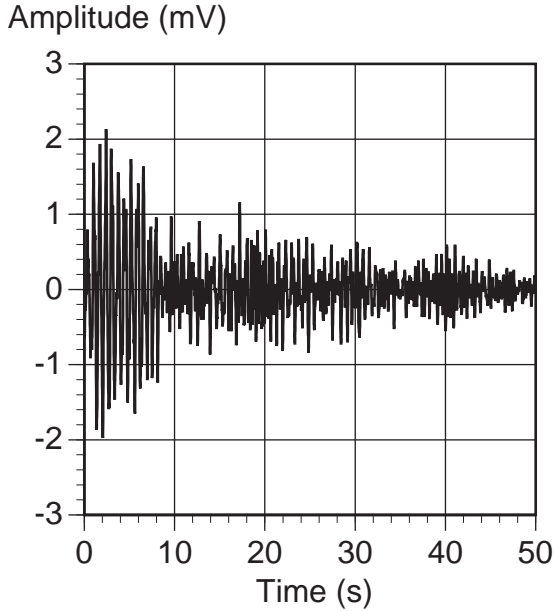} \caption{Seismic plot recorded at 
		the Pontisette station.  Time starts at 00:36:37.3 UT. Further 
		plots of this type can be found in Cevolani et al.  
		(\cite{LUGO2}).}
		\label{pont}
\end{figure}

Observing the seismic plots (e.g.~Fig.\ref{pont}), we can conclude 
that there was a single explosion (for a comparison with nuclear explosions, 
see Pierce et al.  \cite{NUCLEAR}). There is no evidence of multiple explosions,
as it should occur during multiple fragmentation.
Thus, for the Lugo bolide, Eq.~(\ref{e:velo}) can 
be used by assuming that the first fragmentation corresponded to the 
airburst.
 
In order to calculate the flight path angle, we have to solve two 
equations:

\begin{equation}
		\frac{dh}{dt}=V\cdot \sin \theta\, ,
		\label{e:path1}
\end{equation}
 
\begin{equation}
		\frac{d\theta}{dt}=-\frac{\cos 
		\theta}{V}\left(g-\frac{V^{2}}{R+h}\right)\, ,
		\label{e:path2}
\end{equation}

\noindent where $g$ is the gravity acceleration [m/s$^{2}$], $R$ is 
the Earth's radius (we assume $R=6367$ km, for about $45\degr$ 
latitude), and $\theta$ is the flight path angle, measured from the 
horizontal.  We assume that the meteoroid lift can be neglected.  For 
the Tunguska cosmic body, Chyba et al.  (\cite{TUNGUSKA2}) assumed a 
lift value of $10^{-3}$ and found that its influence on the results of 
these calculations is only about $1\%$.  With all these assumptions,
we obtain that the flight path angle during the final part of the 
atmospheric trajectory was $\theta=5.0\degr\pm 0.3\degr$.  Again, we 
have some disagreement with the previous results ($8\degr\div 
20\degr$).  This is probably due to the uncertainty of visual 
observations in these conditions: for such an event the surprise can 
reduce significantly the skills and reliability of eyewitnesses.
 
\section{The reanalysis: Explosion energy}
To obtain an estimate of the explosion energy, we can use the 
relationship for the maximum velocity of displacement of the solid 
rocks, obtained from studies on underground nuclear explosions 
(Adushkin and Nemchinov \cite{IMPACT}).  We rearrange their equation 
in order to calculate the energy, when the distance and the 
displacement velocity are known:

\begin{equation}
		E = k\cdot D^{3}\left(\frac{v}{240}\right)^{12/7}\, ,
		\label{e:kton}
\end{equation}

\noindent where $E$ is the explosion energy in kton of TNT; $D$ is the 
distance of the sensor from explosion [km]; $v$ is the displacement 
velocity [mm/s].  This formula is valid for $D<100$ km: in our case, 
seismic stations were located at distances smaller than 70 km.  The 
coupling coefficient $k$ is introduced to take into account that, in 
order to produce rock displacements, an airburst is less effective 
than an underground nuclear explosion (at least by a factor 100).  
Moreover, there is a difference in the effective energy, because the 
explosion of a meteoroid in the atmosphere does not involve nuclear 
fission, and this contributes about another factor 10.  Finally, there 
is some increase of the wave amplitude with the height of burst up to 
40~km (Pierce et al.  \cite{NUCLEAR}), which typically exceeds a 
factor 2; we assume a power increase by a factor 5.  Overall, we 
estimate $k=100\cdot 10\cdot \frac{1}{5}=200$.
 
\begin{figure}[t]
		\centering
		\includegraphics{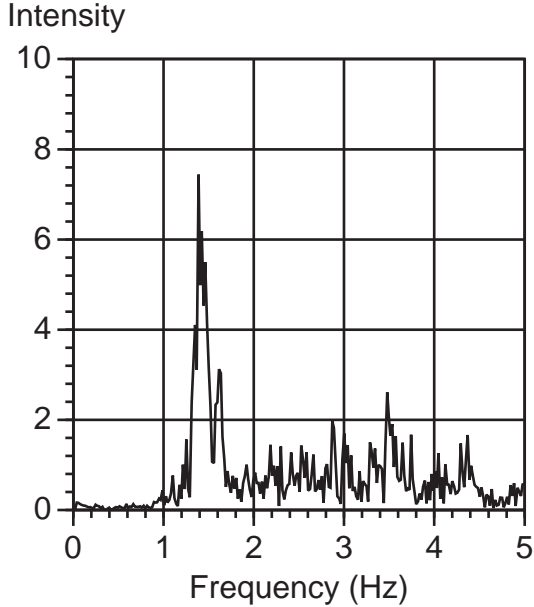}
		\caption{Fourier analysis of the Pontisette seismic plot.}
		\label{pont-f}
\end{figure}

\begin{figure}[t]
		\centering
		\includegraphics{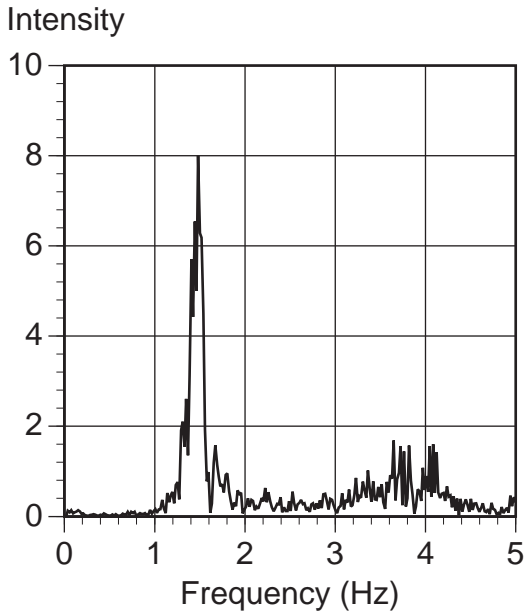} \caption{Fourier analysis of the 
		C\`{a} Fornasina seismic plot.}
		\label{forn-f}
\end{figure}

We have data from six seismic stations (for a complete set of plots 
and other information, see Cevolani et al.  \cite{LUGO2}), but 
transfer functions are available only for the three stations belonging 
to the Microseismic Network of Ferrara.  We have performed a Fourier 
analysis of the waveform and found a peak at 1.4~Hz, for both 
Pontisette and C\`{a} Fornasina, corresponding to the airburst (see 
Figs.~\ref{pont-f} and \ref{forn-f}).  We have not taken into account 
data from the Fiorile d'Albero station, because they show a strong 
background noise overlapping the shock wave and preventing a reliable 
Fourier analysis.
 
The transfer function has a nominal value of 175 mV$\cdot$s/mm for all 
stations and for frequencies greater than 2~Hz.  Below the cutoff 
frequency, the transfer function is drastically reduced, down to a 
value of 10 mV$\cdot$s/mm for 0.5~Hz.  For a frequency of 1.4~Hz, we 
have a transduction factor of 52 mV$\cdot$s/mm.  The final results of 
our calculations for the explosion energy from the seismic data with 
Eq.~(\ref{e:kton}) are shown in Table~\ref{energy}.
 
\begin{table}[h]
		\centering
		\caption{Explosion energy calculated from the seismic data.}
		\begin{tabular}{|l|c|c|c|}
				\hline
				Station & $D$ [km] & $v$ [$\mu$m/s] & $E$ [kton]\\
				\hline
				Pontisette & $59\pm 3$ & $41.040\pm 0.002$ & $14\pm 
				2$\\
				\hline
				C\`{a} Fornasina & $63\pm 3$ & $35.369\pm 0.002$ & 
				$13\pm 2$\\
				\hline
		\end{tabular}
		\label{energy}
\end{table}
 
We consider a mean value of $14\pm 2$~kton, that is $(5.9\pm 
0.8)\times 10^{13}$~J. It is worth noting that we might have obtained 
more accurate values, but the saturation of the Barisano sensor 
introduced an error of $9\%$ in the burst height calculations
(Cevolani et al.  \cite{LUGO2}), which propagates to our results.  On 
the other hand, had we not considered the Barisano data, the available 
data would have been insufficient for a meaningful analysis.
 
When a cometary body or a carbonaceous chondrite enters the 
atmosphere, almost all the kinetic energy is released in the 
explosion.  Then we can calculate the meteoroid mass, taking into 
account that during the path preceding the explosion the cosmic body 
undergoes a limited mass loss:

\begin{equation}
		m = \frac{2E}{V^{2}}=(4\pm 1)\cdot 10^{5} \ [\mathrm{kg}]\, .
		\label{e:mass}
\end{equation}

In order to calculate the visual magnitude of the airburst, we have to 
solve the equation:

\begin{equation}
		L = -\tau \frac{dm}{dt}\frac{V^{2}}{2}\, ,
		\label{e:luminosity}
\end{equation}

\noindent where $\tau$ is the dimensionless coefficient for the meteor 
luminous efficiency.  This coefficient mainly depends on the meteoroid 
speed and is quite uncertain (Ceplecha and McCrosky \cite{PE}).  Some 
authors think that for very bright bolides $\tau$ ranges from $10$ to 
$30\%$ (Brown et al.  \cite{STROBERTS}, McCord et al.,
\cite{MARSHALL1}).  Others assume $\tau$ values between $1.5$ and 
$6.1\%$ (Borovi\v{c}ka and Spurn\'{y} \cite{SL9}, Ceplecha
\cite{TAU}).  Here we assume $\tau = 4.5\%$.
 
Moreover, we assume that the meteoroid dissipated almost all of its 
energy within a scale height.  Then, solving Eq.~(\ref{e:path1}) for 
the time during which the meteoroid exploded, we obtain $t=5.1\pm 
0.8$~s.  The corresponding value for the airburst luminosity is $(5\pm 
1)\cdot 10^{11}$ J/s.  In order to express the luminosity in terms of 
absolute magnitude (i.e., the magnitude as observed at a 100~km 
distance), we can use the equation:

\begin{equation}
		M = -2.5\cdot (\log_{10} L - 2.63)\, ,
		\label{e:magn}
\end{equation}

\noindent where we have rearranged the classical relationship in order 
to use the SI unit system.  From Eq.~(\ref{e:magn}) we obtain $M = 
-22.7\pm 0.5$, a value consistent with visual observations ($-22\div 
-25$).  We stress the importance of the coefficient $\tau$: assuming a 
value of $10\%$,
as suggested by McCord et al.(\cite{MARSHALL1}), we would obtain 
$M\simeq -24$.
 
\section{Further results and discussion}
It is also interesting to check how the results are sensitive to the 
assumed value of the strength $S$.  If we take $S=10^{6}$~Pa, that is 
typical for cometary bodies, we end up with a cosmic body with a speed 
of about 6~km/s and an inclination of $2\degr$.  The mass would be 
about $3\times 10^{6}$~kg and the absolute visual magnitude $-21$.  
The airburst would have been 31~s long.  These values appear unlikely.  
Note that a final velocity of 6~km/s is very close to 4~km/s, which 
Ceplecha (\cite{CEPLECHA}) indicated as necessary to have a meteorite 
fall.  But for Lugo no meteorite was recovered.

We can summarize some features of the Lugo bolide: it had a grazing 
trajectory in the atmosphere, it was probably a carbonaceous 
chondrite, but it exploded at a height higher than usual and with a 
single airburst, without fragmentations.  The recent discovery by the 
NEAR probe of a carbonaceous asteroid (253 Mathilde) with a very low 
density (about $1300$~kg/m$^{3}$) suggests the existence of porous 
bodies (i.e. bodies with internal cavities) among asteroids 
(Yeomans et al.  \cite{MATHILDE}).  If we 
assume that the Lugo bolide was a porous carbonaceous chondrite, we 
have a body which was probably stronger than a cometary fragment, but 
which could explode at a higher altitude than those typical for stony 
objects, because of its porosity.  It is very likely that porosity 
increases the burst efficiency: when ablation removes the surface of 
the body, cavities may appear which increase the aerobraking and 
generate a sudden deceleration.  The kinetic energy then is rapidly 
transformed into heat, so that the body bursts within a scale height.  
This is consistent with a single explosion, without multiple fragmentation, as 
indicated by seismic plots (see Fig.~\ref{pont}).
 
\section{Conclusions}
The Lugo bolide has been reanalysed by taking into account only the 
data recorded by seismic stations.  We summarize the main inferred 
properties of the bolide in the following Table~\ref{summary}.
 
\begin{table}[h]
		\centering
		\caption{Summary on the properties of the Lugo bolide.}
		\begin{tabular}{|l|r|}
				\hline
				Apparition time (UT) & 1993 01 19 00:33:29 $\pm 1$ s\\
				\hline
				Latitude of airburst$^{\mathrm{a}}$ & $44.48\degr \pm 
				0.01\degr$ N\\
				\hline
				Longitude of airburst$^{\mathrm{a}}$ & $11.91\degr \pm 
				0.01\degr$ E\\
				\hline
				Airburst height$^{\mathrm{a}}$ & $30\pm 3$ km\\
				\hline
				Explosion Energy & $14\pm 2$ kton\\
				\hline
				Mass & $(4\pm 1)\cdot 10^{5}$ kg\\
				\hline
				Abs.  Visual Magnitude & $-22.7\pm 0.5$\\
				\hline
				Velocity & $18\pm 3$ km/s\\
				\hline
				Inclination$^{\mathrm{b}}$ & $5.0\degr \pm 0.3\degr$\\
				\hline
				Path azimuth$^{\mathrm{a,c}}$ & $146.5\degr \pm 
				0.5\degr$\\
				\hline
		\end{tabular}
		\label{summary}
		\begin{list}{}{}
		\item[$^{\mathrm{a}}$] Calculated in Cevolani et al.  
		(\cite{LUGO2}).  \item[$^{\mathrm{b}}$] Over the horizon.  
		\item[$^{\mathrm{c}}$] Clockwise from North.
		\end{list}
\end{table}
 
We are now carrying out calculations on the orbit and the dynamical 
evolution this bolide, whose results will be available soon.  However, 
from the analysis described here it appears likely that the meteoroid 
was a porous carbonaceous chondrite, somehow similar in constitution 
to the asteroid 253 Mathilde.  The porosity would have increased the 
braking and as a consequence the airburst occurred at a height higher 
than for a compact carbonaceous chondrite object.
 
\begin{acknowledgements}
I wish to thank P.~Farinella who drew my attention to the special 
properties of the asteroid 253~Mathilde and for constructive review. 
I wish also to thank T.J.~Jopek and Z.~Ceplecha, 
for valuable discussions concerning the velocity calculations, and a 
referee, M.P.~Goda, for useful comments.
\end{acknowledgements}

\end{document}